\documentstyle[aps,prl,xspace]{revtex}

\begin{document}

\newcommand{\kn}{${\rm K}^0$\xspace}
\newcommand{\knt}{$\widetilde{{\rm K}^0}$\xspace}

\newcommand{\nri}{$\nu_{\rm I}$\xspace}
\newcommand{\nrii}{$\nu_{\rm II}$\xspace}
\newcommand{\nrit}{$\tilde{\nu}_{\rm I}$\xspace}
\newcommand{\nriit}{$\tilde{\nu}_{\rm II}$\xspace}


\title{Two Majorana Neutrino Mass Doublets with Thorough Maximal Doublet 
Mixing from an Analogy with the \kn-Meson Oscillations}

\author{E.M. Lipmanov\footnote{e-mail: elipmanov@yahoo.com}}

\address{40 Wallingford Rd. \#272, Brighton, MA 02135, USA}

\maketitle
 
\begin{abstract}
  In the recent note~\cite{lipmanov} we argued that most of the
  available positive and negative neutrino oscillation data can be
  incorporated in a four-neutrino weak interaction model with one
  physical condition: thorough maximal mixing of the neutrino components
  in each of the two mass doublets.  On the level of CP-invariance an
  extended Pontecorvo analogy between the neutrino oscillations and the
  \kn-meson oscillations, which implied maximal violation of the lepton
  charge conservation in the neutrino mass matrix exclusively, can
  afford a physical clue to the three data dictated assumptions in a
  model with three lepton flavors and only two 4-component neutrino
  generations: four phenomenological neutrinos with two neutrino mass
  doublets, one sterile neutrino in the oscillations, and thorough
  maximal neutrino doublet mixing.  This analogy predicts that all the
  four neutrino mass eigenstates are of the Majorana type.
\end{abstract}


\pacs{14.60. st + pg, 12.15 fg.}

\keywords{Nu-K0 oscillation analogy; Majorana neutrinos}


In ref.~\cite{lipmanov} a simple explicit neutrino mixing and
oscillation model with four phenomenological neutrinos~\cite{numodels}
-- the three weak interaction eigenstates $\nu_e, \nu_\mu$, and
$\nu_\tau$ plus one sterile neutrino $\nu_s$ -- is built with one
condition: the neutrino mass eigenstates enter the four phenomenological
neutrinos only in the form of the eigenstates of the mass doublet
neutrino exchange symmetry, symmetric or anti-symmetric.  At the level
of CP-invariance of the weak interactions this condition is equivalent
to the assumption of a thorough maximal mixing of the neutrino mass
eigenstates in each of the neutrino mass doublets.  The model predicts
naturally large oscillation amplitudes for the atmospheric and solar
neutrino anomalies and is in good agreement with the positive
Super-Kamiokande data~\cite{superk} if the LSND data~\cite{lsnd} are
accepted.  It is also in agreement with the negative reactor and
accelerator data.  Though there is yet a long way ahead to the
experimental verification of this model, it is certainly interesting
that it gets a strong physical support from an extended Pontecorvo
analogy between the neutrino oscillations and the well known \kn-meson
oscillations, where the ``particle'' oscillations had been discovered
first.  As a useful guiding suggestion, this interesting physical
analogy is considered below from the standpoint of a possible clue to
the thorough maximal neutrino doublet mixing.

Vacuum neutrino oscillations of the type nu-anti-nu in analogy with the
\kn-\knt oscillations were considered in the original Pontecorvo
paper~\cite{pontecorvo}.  Accordingly, we assume here that the lepton
charge is conserved in the weak interactions (as strangeness of 
the \kn and \knt-mesons in the strong interactions), but not in the
neutrino mass matrix.    Its eigenstates can then be the truly neutral
(Majorana) neutrino mass states (as the \kn$\!\!\!_{1,2}$ mesons).  In the
unreal case with only two lepton flavors $e$ and $\mu$, the minimal
model is bounded to one 4-component neutrino $\nu$ and the definition of 
the two flavor neutrinos is
\begin{equation}
  \label{eq:linear_comb}
  \nu_e = \nu = \frac{1}{\sqrt{2}}(\nu_1+\nu_2),~~\nu_\mu = \tilde{\nu} =
  \frac{1}{\sqrt{2}}(\nu_1-\nu_2), 
\end{equation}
\noindent with $\nu_1$ and $\nu_2$ being the two neutrino Majorana mass
eigenstates,
\begin{equation}
  \label{eq:majorana_states}
  \nu_1 = \frac{1}{\sqrt{2}}(\nu+\tilde{\nu}),~~\nu_2 =
  \frac{1}{\sqrt{2}}(\nu-\tilde{\nu}), 
\end{equation}
\noindent
in strict analogy with the definition of the \kn$\!\!_1$ and \kn$\!\!_2$
meson states with definite combined CP-parities.  All four neutrino
states participate in the weak interactions here, and there is no
sterile neutrino.  The definition~(\ref{eq:linear_comb}) means that the
electron and muon neutrinos have opposite lepton charges, but it does
not preclude the strict conservation of the lepton charge in the weak
interactions.  The transitions $(\nu_e)_L \leftrightarrow 
(\tilde{\nu}_\mu)_R$ and $(\nu_\mu)_L \leftrightarrow (\tilde{\nu}_e)_R$ 
are induced by the neutrino
masses, and their probabilities are very small for relativistic
neutrinos.  The only way to the experimental detection of the truly
neutral neutrino constituent states $\nu_1$ and $\nu_2$ within the weak
interaction eigenstates in Eq.~(\ref{eq:linear_comb}) is through the
observation of the neutrino oscillations $(\nu_e \leftrightarrow
\nu_\mu; ~\nu_e \rightarrow \nu_e; ~\nu_\mu \rightarrow
\nu_\mu)$ with maximal oscillation amplitudes -- in contrast to the
\kn-meson case, where the \kn$\!\!_1$ and \kn$\!\!_2$ constituents of
the kaons had been discovered first by their decay modes, and the
oscillations had been observed later.  In the realistic case with three
lepton flavors $e, \mu$ and $\tau$, the minimal model is bounded to two
4-component neutrinos \nri~and \nrii.  The necessity of a second
4-component neutrino with the same lepton charge is a data dictated
condition -- by the lepton flavor proliferation and the LSND indication.
The most general mixing pattern of these two neutrinos and their
anti-particles \nrit~and \nriit, with the substantial condition of
strict conservation of the lepton charge in the short-range weak
interactions, is a $4\times4$ unitary mixing matrix with two parameters
which does not mix the neutrino particle and anti-particle states.  It
determines the following four 4-component phenomenological neutrino
states:
\begin{eqnarray}
  \label{eq:nue}
  \nu_e & = & ~~\nu_{\rm I} \cos \theta + \nu_{\rm II} \sin \theta,\\
  \label{eq:numu}
  \nu_\mu & = & -\nu_{\rm I} \sin \theta + \nu_{\rm II} \cos \theta,\\
  \label{eq:nutau}
  \nu_\tau & = & ~~\tilde{\nu}_{\rm I} \sin \phi + \tilde{\nu}_{\rm II} \cos \phi,\\
  \label{eq:nus}
  \nu_s & = & ~~\tilde{\nu}_{\rm I} \cos \phi - \tilde{\nu}_{\rm II} \sin \phi.
\end{eqnarray}
\noindent
The weak interaction eigenstates are $(\nu_e)_L$, $(\nu_\mu)_L$ and 
$(\nu_\tau)_L$; the appearance of one 2-component sterile neutrino 
$(\nu_s)_L$ is an inevitable result here.  Only eight of the sixteen 
neutrino components
in the Eqs.~(\ref{eq:nue})--(\ref{eq:nus}) are independent: the right
neutrinos $(\nu_e)_R$ and $(\nu_\mu)_R$ (and the corresponding left
anti-neutrinos) are linear superpositions of the right anti-neutrinos
$(\tilde{\nu}_\tau)_R$ and $(\tilde{\nu}_s)_R$ (and the corresponding
left neutrinos), and vice-versa (there are only two independent 
4-component neutrino generations, e.g. $\nu_e$ and $\nu_\mu$ and in the
particular case with $\phi = -\theta$ it would be $\nu_\tau =
\tilde{\nu_\mu}, \nu_s = \tilde{\nu_e}$).  
It means that the three lepton numbers 
$n_e$, $n_\mu$, and $n_\tau$ are not conserved in the local weak
interactions if we take into account the effects of the neutrino
masses.  Only one lepton charge can be strictly conserved in the local
weak interactions of the massive neutrinos, it is the one carried by the
4-component neutrinos \nri and \nrii.  The condition in the
Eqs.~(\ref{eq:nue}) and (\ref{eq:numu}) that the $\nu_e$ and $\nu_\mu$
share the same two neutrino states is an indication which follows from
the LSND data~\cite{lsnd}.  The formulation of the neutrino mixing
model~(\ref{eq:nue})--(\ref{eq:nus}) with two 4-component neutrinos is
unique because of its apparent discrete exchange symmetry $\nu_{\rm I}
\leftrightarrow \tilde{\nu}_{\rm I},~\nu_{\rm II} \leftrightarrow
\tilde{\nu}_{\rm II}$ and the above mentioned condition of strict
conservation of the lepton charge in the weak interactions, plus the
indication from the LSND effect.  The only two free parameters here, the
two mixing angles $\theta$ and $\phi$, have to be determined from the
experimental data (the parameter $\theta$ is the LSND mixing angle).
The neutrino pairs $(\nu_e,\nu_\mu)$ and $(\nu_\tau,\nu_s)$ have
opposite lepton charges with a definition of the ``leptons'':
$\nu_e,\nu_\mu,\tilde{\nu}_\tau,\tilde{\nu}_s,e^-,\mu^-$, and $\tau^+$;
and the ``anti-leptons'':
$\tilde{\nu}_e,\tilde{\nu}_\mu,\nu_\tau,\nu_s,e^+,\mu^+$, and $\tau^-$.
The probabilities of the transitions $(\nu_e)_L, (\nu_\mu)_L \leftrightarrow
(\tilde{\nu}_\tau)_R, (\tilde{\nu}_s)_R$ and $(\tilde{\nu}_e)_R, 
(\tilde{\nu}_\mu)_R \leftrightarrow (\nu_\tau)_L, (\nu_s)_L$, 
induced by the neutrino masses, are
very small for relativistic neutrinos with $m_\nu^2/E^2_\nu\ll 1$.

Above we considered the two 4-component neutrinos \nri and \nrii as
regular Dirac particles with masses $m_1$ and $m_2$.  Only at this level
is the lepton charge strictly conserved in the weak interactions of
massive neutrinos.  In accordance with the nu$\sim$\kn analogy we assume
now that the lepton charge is not conserved in the neutrino mass matrix,
and its eigenstates $\nu_i$ and $\nu^\prime_i$, $i$=1,2, are the truly
neutral neutrino mass states:
\begin{eqnarray}
  \label{eq:cp-parity}
    \nu_{\rm I} =
    \frac{1}{\sqrt{2}}(\nu_1+\nu^\prime_1),&~~&\tilde{\nu}_{\rm I} = 
  \frac{1}{\sqrt{2}}(\nu_1-\nu^\prime_1),\nonumber \\
    \nu_{\rm II} =
    \frac{1}{\sqrt{2}}(\nu_2+\nu^\prime_2),&~~&\tilde{\nu}_{\rm II} = 
  \frac{1}{\sqrt{2}}(\nu_2-\nu^\prime_2).
\end{eqnarray}
\noindent
With the statement in Eq.~(\ref{eq:cp-parity}) the neutrino mixing
model~(\ref{eq:nue})--(\ref{eq:nus}) has the special feature of
``thorough maximal neutrino doublet mixing'' which is discussed in
ref.~\cite{lipmanov}; the correspondence with the notations in this
reference is $\nu^s_{1,2} = \nu_{\rm I,II},~\nu^a_{1,2} =
\tilde{\nu}_{\rm I,II}$, i.e. the two 4-component neutrinos $\nu_{\rm
  I,II}$ and their anti-particles $\tilde{\nu}_{\rm I,II}$ in the
Eqs.~(\ref{eq:nue})--(\ref{eq:nus}) are the four eigenstates $\nu^s_i$
and $\nu^a_i$, $i$=1,2, of the exchange symmetry operators of the
neutrino Majorana mass eigenstates $\nu_i$ and $\nu^\prime_i$ in the two
mass doublets.  Thus we can suggest that the ``thorough maximal neutrino
doublet mixing'' of ref.~\cite{lipmanov} is due to the physical
conditions of lepton charge conservation in the weak interactions and
that the neutrino mass eigenstates are of the Majorana type, with
definite combined CP-parities; the thorough maximal neutrino doublet
mixing in the weak interactions is a manifestation of the implied
maximal violation of the lepton charge conservation in the neutrino mass
matrix, which is the sole effective location of lepton nonconservation
in the present model.  Note that the implied here characteristics of the
neutrino mass matrix with regard to the discrete symmetries are opposite
to the ones of the neutrino weak interactions: the neutrino mass matrix
violates maximally the lepton charge conservation, but does not mix the
components of the two neutrino generations \nri and \nrii.  Therefore,
the lepton charge nonconservation in the weak interactions of the
massive Majorana neutrinos is induced only by the neutrino mass doublet
splittings and should be very small in comparison with the
aforementioned lepton number nonconservations in the important case of
very narrow neutrino mass doublets.  Since no neutrino decay modes are
known, the only way to the experimental detection of the Majorana mass
constituent states in the weak interaction neutrino eigenstates is by
the observations of the neutrino oscillations with maximal amplitudes.
Thus the discovery of large neutrino mixing in the Super-Kamiokande
oscillation experiment~\cite{superk} can be regarded as an indication in
this direction.

The thorough maximal neutrino doublet mixing in the
model~(\ref{eq:nue})--(\ref{eq:cp-parity})  leads to a number of
distinct and characteristic predictions~\cite{lipmanov}.  The data of
the solar, atmospheric and the LSND experiments can be explained here
with the following values of the neutrino mass squared
differences~\cite{numodels},
\begin{eqnarray}
  \label{eq:dm2}
  \Delta m_1^2 & \equiv & \Delta m^2_{solar} \sim 10^{-10}(Vac),~{\rm
    or} \sim 10^{-5}(MSW)~{\rm eV}^2,\nonumber\\
  \Delta m_2^2 & \equiv & \Delta m^2_{atm} \sim 10^{-3} {\Large -}
  10^{-2}~{\rm eV}^2,\nonumber\\
  \Delta m_{12}^2 & \equiv & \Delta m^2_{LSND} \sim 1~{\rm eV}^2,
\end{eqnarray}
\noindent
and a possible scheme of the neutrino mass spectrum
\begin{equation}
  \label{eq:doublet}
  \underbrace{\overbrace{m_1 < m_1^\prime}^{solar} \ll 
    \overbrace{m_2 < m_2^\prime}^{atm}}_{LSND}~.
\end{equation}
\noindent
In the other possible scheme the positions of the ``solar'' and the
``atm'' mass splittings are exchanged.  The long-baseline oscillations
in the mixing model (\ref{eq:nue})--(\ref{eq:cp-parity}), (\ref{eq:dm2})
and (\ref{eq:doublet}), including the atmospheric and solar ones with
maximal oscillation amplitudes, are lepton charge neutrino-antineutrino
$\nu_{\rm I,II}\leftrightarrow\tilde{\nu}_{\rm I,II}$ oscillations (as
the oscillations of strangeness in the $K^0$ case), whereas the
short-baseline ones are neutrino generation $\nu_{\rm
  I}\leftrightarrow\nu_{\rm II}$ and $\tilde{\nu}_{\rm
  I}\leftrightarrow\tilde{\nu}_{\rm II}$ oscillations (with no $K^0$
analogy: they would remain unchanged even if the neutrino mass matrix
were of the regular Dirac type with no neutrino doublet splittings).
The equations below are explicitly written in terms of the first scheme,
the conclusions are valid for both of them.  The probability of the
appearance $\nu_\mu \leftrightarrow \nu_e$ oscillations:
\begin{equation}
  \label{eq:lsnd_prob}
  \left | \left \langle \nu_\mu(0) | \nu_e(L) \right \rangle \right |^2
  = \sin^22\theta \left [ \left \langle \! \! \! \left \langle \sin^2 
        \left ( \frac{\Delta m^2_{12}L}{4E}
        \right ) \right \rangle \! \! \!
    \right \rangle - \frac{1}{4}\sin^2\left ( \frac{\Delta m_1^2 L}{4E} 
    \right ) - \frac{1}{4}\sin^2\left ( \frac{\Delta m_2^2 L}{4E} 
    \right ) \right ],
\end{equation}
\noindent
where $\theta = \theta_{LSND}$, and the symbol $\langle \! \langle ~
\rangle \! \rangle$ in the first term denotes the arithmetic mean value
of the appropriate four factors related to the four ``large'' mass
squared differences among the two neutrino mass doublets.  The notations
are
\begin{equation}
  \label{eq:notation}
  \Delta m^2_{1,2} = m^2_{1^\prime,2^\prime} - m^2_{1,2}~,\;\; \Delta m^2_{12} 
  \cong m^2_2 - m^2_1~,
\end{equation}
$E$ is the initial beam energy and $L$ is the distance from the source.
From the Eq.~(\ref{eq:lsnd_prob}), the $\nu_\mu \leftrightarrow \nu_e$
oscillations, both the short-baseline and the long-baseline, are
determined by one factor $\sin^22\theta$, which is the LSND oscillation
amplitude with the estimation~\cite{lsnd} $\sin^22\theta \approx 3\times 
10^{-3}$.

The probability of the $\nu_\mu \rightarrow (\nu_\tau + \nu_s)$
appearance oscillations, 
\begin{equation}
  \label{eq:w_app}
  W(\nu_\mu\rightarrow\nu_\tau+\nu_s) =
  \cos^2\theta\sin^2  \left (
    \frac{\Delta m_2^2 L}{4 E} \right ) + \sin^2\theta\sin^2  \left (
    \frac{\Delta m_1^2 L}{4 E} \right ) ,
\end{equation}
\noindent
is independent of the second mixing angle $\phi$ and also of the
neutrino mass doublet separation $\Delta m^2_{12}$ and, therefore,
describes only long-baseline oscillations, atmospheric and possibly
terrestrial.  The first term in this equation is the dominant one, it
describes the main part of the atmospheric $\nu_\mu$ oscillations with
the amplitude
\begin{equation}
  \label{eq:a_atm}
  A_{atm} \cong \cos^2\theta \cong 1.
\end{equation}

The probability of the coming long-baseline accelerator $\nu_\mu$
survival oscillations can be described with a good approximation by the
single term
\begin{equation}
  \label{eq:w_lbl}
  W(\nu_\mu\rightarrow\nu_\mu) \cong 1 -
  W(\nu_\mu\rightarrow\nu_\tau+\nu_s) \cong  
  \cos^2 \left (
    \frac{\Delta m_2^2 L}{4 E} \right ) ,
\end{equation}
\noindent
with L the distance from the $\nu_\mu$ source to the detector.

The probability of the $\nu_e \rightarrow (\nu_\tau + \nu_s)$ appearance 
oscillations is:
\begin{equation}
  \label{eq:w_app_e}
  W(\nu_e\rightarrow\nu_\tau+\nu_s) =
  \cos^2\theta\sin^2  \left (
    \frac{\Delta m_1^2 L}{4 E} \right ) + \sin^2\theta\sin^2  \left (
    \frac{\Delta m_2^2 L}{4 E} \right ).
\end{equation}
\noindent
The first term in this equation is the dominant one and describes the
solar neutrino appearance oscillations with the amplitude
\begin{equation}
  \label{eq:a_solar}
  A_{solar} \cong \cos^2\theta \cong 1.
\end{equation}
\noindent
The probability of the solar vacuum neutrino survival oscillations,
\begin{equation}
  \label{eq:w_vac}
  W(\nu_e\rightarrow\nu_e) \cong 1 -
  W(\nu_e\rightarrow\nu_\tau+\nu_s) \cong  
  \cos^2 \left (
    \frac{\Delta m_1^2 L_I}{4 E} \right ) ,
\end{equation}
\noindent
is just the original still viable Pontecorvo~\cite{pontecorvo} 2-neutrino 
maximal mixing solution for the solar neutrino deficit (here $L_I$ is
the Earth-Sun distance).

The probability of the $\nu_\tau\rightarrow(\nu_\mu+\nu_e)$ appearance
oscillations,
\begin{equation}
  \label{eq:w_tau}
  W(\nu_\tau\rightarrow\nu_\mu+\nu_e) =
  \cos^2\phi\sin^2  \left (
    \frac{\Delta m_2^2 L}{4 E} \right ) + \sin^2\phi\sin^2  \left (
    \frac{\Delta m_1^2 L}{4 E} \right ) ,
\end{equation}
is independent of the first mixing angle $\theta$, and of
$\Delta m^2_{12}$, and suggests that there are only long-baseline
oscillations of this kind.

The probability of the $\nu_\tau\rightarrow\nu_s$ oscillations is
described by the expression at the right side of the
Eq.~(\ref{eq:lsnd_prob}) with $\theta\rightarrow\phi$.  From the
big-bang nucleosynthesis implications~\cite{bigbang} probobly it follows
\begin{equation}
\label{eq:nucleosynthesis}
\sin^22\phi< 10^{-7}.
\end{equation}
\noindent Such a constraint implies (even if the astrophysical
estimations would weaken it by several orders of magnitude) that from
the continuum of possibilities in the atmospheric
$\nu_\mu\rightarrow(\nu_\tau+\nu_s)$ and solar
$\nu_e\rightarrow(\nu_\tau+\nu_s)$ oscillation dominances only the two
extremes $(\nu_\mu\rightarrow\nu_\tau;\nu_e\rightarrow\nu_s)$ or
$(\nu_\mu\rightarrow\nu_s;\nu_e\rightarrow\nu_\tau)$ are allowed.

The main conclusions are: 1). In the present model, the basic
description of the neutrinos in the weak interactions is by charge
carrying 4-component spinors as in all the other known cases of the
fermions.  The peculiarity of the neutrino case is that here it is
rather enough to have only two 4-component neutrino generations, instead
of three 4-component generations used for the charged leptons and
quarks.  These two neutrino generations are reduced to four 2-component
Majorana neutrino mass eigenstates because of the maximal parity
nonconservation in the weak interactions and the maximal lepton charge
nonconservation in the neutrino mass matrix.  The complete superposition
set of the left elements of these four Majorana neutrinos is structuring
the complete set of the left components of the four phenomenological
neutrinos $(\nu_e)_L , (\nu_\mu)_L, (\nu_\tau)_L$ and $(\nu_s)_L$ in the
Eqs.~(\ref{eq:nue})--(\ref{eq:nus}), and this structure determines the
different types of neutrino oscillations; 2). The atmospheric $\nu_\mu$
and the solar $\nu_e$ oscillation amplitudes are naturally large if the
LSND data are accepted; 3). The transitions
$\nu_\mu\rightarrow\nu_\tau+\nu_s$ and $\nu_e\rightarrow\nu_\tau+\nu_s$
give the main contributions to the atmospheric and the solar neutrino
oscillations because of the small LSND mixing angle $\theta$.  The
dependence on the second mixing angle $\phi$ leads to the different
``complementarity'' relations between the separate $\nu_\tau$ and
$\nu_s$ channels in the sum $(\nu_\tau+\nu_s)$; 4). The contribution of
the $\nu_\mu \leftrightarrow \nu_e$ transitions to both of these
oscillations are smaller by more than two orders of magnitude.  The last
three inferences are in good agreement with the positive
Super-Kamiokande data~\cite{superk}; 5). The four phenomenological
neutrinos are divided into two different pairs $(\nu_e,\nu_\mu)$ and
$(\nu_\tau,\nu_s)$ in conformity with the two different pairs of the
particle and anti-particle states of the 4-component neutrinos
(\nri,\nrii) and (\nrit,\nriit).  This feature of the model leads to the
prediction of negative results for the $\nu_\mu,\nu_e \rightarrow
\nu_\tau,\nu_s$ oscillation searches in the short-baseline accelerator
experiments such as CHORUS and NOMAD~\cite{chorus-nomad}, and of large
oscillation amplitudes at the appropriate long-baseline experiments, in
contrast to the $\nu_\mu\leftrightarrow\nu_e$ and, probably,
$\nu_\tau\leftrightarrow\nu_s$ oscillations; 6). The deviation from
unity of the $\nu_e\rightarrow\nu_e$ survival probability is small,
$\leq \sin^22\theta_{LSND}$ for oscillation distances $L \ll 2\pi
E/\Delta m^2_1$.  This is in good agreement with the $\nu_e$
disappearance oscillation data such as Bugey~\cite{bugey}, and the
appearance oscillation $\nu_\mu\rightarrow\nu_e$ data from the BNL E776
experiment~\cite{bnl}, and also with the long-baseline reactor $\nu_e$
disappearance CHOOZ experiment~\cite{chooz} where this restriction on
the distances is fulfilled quite well; 7). Though all the neutrino mass
eigenstates are of the Majorana nature, the double $\beta$ decay in the
model is extremely suppressed by the factors of the neutrino doublet
splittings with the dominant $\Delta m^2_1$ from the Eq.~(\ref{eq:dm2}),
because of the condition of lepton charge conservation in the weak
interactions of unsplit doublet neutrinos.  The allowed by the lepton
charge conservation reactions $\mu^- + {\rm A(Z)} \rightarrow {\rm
  A(Z-2)} + \tau^+$ and also decay modes $\tau^+ \rightarrow
\mu^-+\pi^++\pi^+$, etc., are mediated by the neutrino masses, but the
latter introduce small factors.

The predictions above appear to be straightforward inferences of the
analogy between the neutrino and the \kn-meson oscillation phenomena in
the minimal description with only two 4-component neutrino generations
and three lepton flavors.  It is a remarkable fact that this simple and
interesting physical analogy is in good agreement with the majority of
the available positive and negative oscillation data.  Note, however,
that the acceptance of the LSND data~\cite{lsnd} is crucial for this
agreement.  

I would like to thank Alec Habig and Mark Messier for the help with
preparation of the manuscript and interesting conversations.

\end{document}